\documentclass[onecolumn]{elsart3p} 
\usepackage{amsmath}
\usepackage{amssymb}
\usepackage{bm}
\usepackage{epsfig}
\usepackage{float}
\usepackage{amsmath}
\usepackage{amssymb}

\newcounter{bla}

\def\qlim       {\lim_{\qq\rightarrow 0}}

\def\gd         {\delta} 
\def\gee        {\varepsilon}
\def\gl         {\lambda}
\def\go         {\omega}

\def\gr         {\rho}

\def\la         {\langle}
\def\ra         {\rangle}
\def\kk         {{\bf k}}

\def\qq         {{\bf q}}
\def\rr         {{\bf r}}

\def\GG         {{\bf G}}

\renewcommand{\[}{\left[}
\renewcommand{\]}{\right]}
\renewcommand{\(}{\left(}
\renewcommand{\)}{\right)}

\newcommand{\yambo} {{\normalfont\ttfamily yambo}}
\newcommand{\ypp}  {{\normalfont\ttfamily ypp}}
\newcommand{\ay}  {{\normalfont\ttfamily a2y}}
\newcommand{\py}  {{\normalfont\ttfamily p2y}}
\newcommand{\ey}  {{\normalfont\ttfamily e2y}}

\begin{document}
\begin{frontmatter}

\title{\yambo: an \textit{ab initio} tool for excited state calculations}

\author[a,e]{Andrea Marini\thanksref{author}},
\author[b,e]{Conor Hogan},
\author[c,e]{Myrta Gr\"uning}, and
\author[d,e]{Daniele Varsano}

\thanks[author]{Corresponding author}

\address[a]{
 Dipartimento di Fisica, 
 CNISM,
 and SMC Institute for Statistical Mechanics and Complexity,
 Universit\`a di Roma ``Tor Vergata'',
 Via della Ricerca Scientifica 1,
 I--00133 Roma, Italy}

\address[b]{
 Dipartimento di Fisica and INFM--CNR,
 Universit\`a di Roma ``Tor Vergata'',
 Via della Ricerca Scientifica 1,
 I--00133 Roma, Italy}

\address[c]{
Unit\'e PCPM, Universit\'e Catholique de Louvain, 1348 Louvain-la-Neuve, Belgium}
\address[d]{
National Center on nanoStructures and Biosystems at Surfaces (S3)
of INFM-CNR, I--41100 Modena, Italy}
\address[e]{European Theoretical Spectroscopy Facility (ETSF)} 

\begin{abstract}
\yambo\ is an {\it ab initio} code for calculating  quasiparticle
energies and optical properties of electronic systems 
within the framework of many-body perturbation theory 
and time-dependent density functional theory. 
Quasiparticle energies are calculated within the $GW$ 
approximation for the self-energy.
Optical properties are evaluated either by solving the Bethe--Salpeter equation or by using
the adiabatic local density approximation.
\yambo\ is a plane-wave code that, although 
particularly suited for calculations of 
periodic bulk systems, has been applied to a large variety of
physical systems.
\yambo\ relies on efficient numerical techniques devised to treat 
systems with reduced dimensionality, or with a large number of
degrees of freedom.
The code has a user-friendly command-line based
interface, flexible I/O procedures and is interfaced to several
publicly available density functional ground-state codes.
\begin{flushleft}
%
%
\end{flushleft}


\end{abstract}

\end{frontmatter}

\noindent{\bf PROGRAM SUMMARY}\\

\begin{small}
\noindent
{\em Program title:}  \yambo                                     \\\\
{\em Journal Reference:}                                      \\\\
{\em Catalogue identifier:}                                   \\\\
{\em Program obtainable from:} http://www.yambo-code.org       \\\\
{\em Licensing provisions:} This program is distributed under the GNU
  General Public License v2.0 ( see http://www.gnu.org/ for details)   \\\\
{\em Programming language:} Fortran 95, C  \\\\
{\em Computer(s) for which the program has been designed:} any computer architecture, running any 
flavor of UNIX \\\\
{\em Operating system(s) under which the program has been tested:} GNU/Linux, AIX, Irix, OS/X \\\\
{\em RAM required to execute program with typical data:} 10--1000 Mbytes        \\\\
{\em Number of processors used:} Up to 100                         \\\\
{\em Supplementary material:} Manuals, tutorials, background theory and sample input files are available for download or inspection from the website http://www.yambo-code.org                                
{\em Classification:} 7.3  Electronic Structure, 4.4  Feynman Diagrams, 7.2  Electron Spectroscopies,
  18 Optics \\ \\
{\em External routines/libraries:} 
  \begin{itemize} 
  \item[$\cdot$] BLAS (http://www.netlib.org/blas/)
  \item[$\cdot$] LAPACK (http://www.netlib.org/lapack/)
  \item[$\cdot$] MPI (www-unix.mcs.anl.gov/mpi/) is optional.
  \item[$\cdot$] BLACS (http://www.netlib.org/scalapack/) is optional.
  \item[$\cdot$] SCALAPACK (http://www.netlib.org/scalapack/) is optional.
  \item[$\cdot$] FFTW (http://www.fftw.org/) is optional.
  \item[$\cdot$] netCDF (www.unidata.ucar.edu/software/netcdf/) is optional.
  \end{itemize}  
%
{\em Nature of problem:}\\
   Calculation of excited state properties (quasiparticles, excitons, plasmons)
   from first principles. 
   \\\\
{\em Solution method:}\\
   Many body perturbation theory (Dyson equation, Bethe--Salpeter equation) and 
   time-dependent density functional theory. 
   Quasiparticle approximation. Plasmon-pole model for the dielectric screening. 
   Plane--wave basis set with norm conserving pseudopotentials.
   \\\\
 {\em Unusual features:}\\
 During execution, \yambo\ supplies estimates of the elapsed and remaining 
 time for completion of each runlevel. Very friendly shell-based user-interface. 
   \\\\
{\em Additional comments:}\\
 \yambo\ was known as ``SELF'' prior to GPL release. It belongs to the suite of codes 
 maintained and used by the European Theoretical Spectroscopy Facility\,(ETSF)~\cite{etsf}
   \\\\
{\em Running time:}\\
 The typical \yambo\, running time can range from a few minutes to some days depending
 on the chosen level of approximation, and on the property and physical system under study.
   \\\\
\end{small}
 
\hspace{1pc}
\noindent\textbf{LONG WRITE-UP}

\section{Introduction}

{\it Ab initio} calculations in the framework of density functional theory\,(DFT)~\cite{DFT_review}
have yielded high-quality results for a large variety of systems, ranging from
periodic solids to molecules and nanostructures~\cite{DFT_review}.
These results are however mostly limited to
quantities related to the electronic {\it ground state}, whereas additional
phenomena that occur in the excited state are not correctly described~\cite{rmp}.

It was recognized at an early stage~\cite{dft_gap_problem} that in extended systems the 
standard approximations for DFT --\, the local density\,(LDA) or 
generalized gradient\,(GGA) approximations\,-- 
fail to describe,  among other effects, the band gap of 
insulators and semiconductors~\cite{rmp}.
In contrast, many-body perturbation theory\,(MBPT)~\cite{MBPT_review} provides, by means of the
quasiparticle\,(QP) concept, a more adequate and accurate approach that yields
band gaps (and band structures, in general) in good agreement with the experimental values~\cite{rmp}.

While the first successful QP calculations were performed in the mid
80's~\cite{hyber}, it was only in the late 90's that many-body effects have
been  included in the {\it ab initio} calculation of optical
properties of real materials~\cite{albrecht}. 
It is now well known that a quantitative
description of the optical response of an interacting electron system 
must account for electron--hole interactions (excitonic
effects)~\cite{rmp}.  
This is achieved
by solving the Bethe--Salpeter\,(BS) equation for the electron--hole
Green's function, within the MBPT framework. Nowadays, solving 
the BS equation is 
the state-of-the-art approach for calculating optical properties
in extended systems.
The importance of including excitonic effects is clear on comparison with
the experimental absorption spectra of semiconductors and insulators.
In particular, for wide-gap 
insulators there is hardly any resemblance between the spectrum calculated within a
noninteracting theory and the experiment. 

An alternative approach to the study of correlation in many-body systems 
is given by time dependent DFT\,(TDDFT)~\cite{TDDFT_review}.
Similar to the paradigm of DFT for ground-state properties, TDDFT has 
emerged as a very powerful tool for the description of excited
states. In principle TDDFT is an exact theory for neutral excited state properties.
Nevertheless, in practice it has a number of commonly cited 
failures related to the approximation
of the exchange-correlation\,(xc) kernel.
One example is the difficulty
encountered when studying the optical properties of
extended systems; another is the severe underestimation of high-lying
excitation energies in molecules.
On the other hand, the combination of  TDDFT with simple approximations for the
xc kernel (based  
on the homogeneous electron gas) has been successfully applied to the study   of the
optical response of molecules and nanostructures. 

Quasiparticles, excitons and plasmons are the
excitations that can be calculated using the \yambo\, code.
These excitations are ubiquitous in the {\it ab initio} description of the 
electronic and optical properties of any physical system.
The \yambo\, code uses as input the result of standard DFT 
calculations obtained by means of publicly available codes~\cite{pwscf,abinit}.
The theoretical tools implemented in \yambo\, are TDDFT and the BS equations for the
response function and the Dyson equation in the $GW$ approximation for the QPs.

The paper is structured as follows. In Section~\ref{sec:theo} we introduce the most important theoretical concepts as they are utilized in \yambo, before describing 
some of the most important 
numerical algorithms implemented in the code in Section~\ref{sec:numerics}.
Section~\ref{sec:overview} outlines the structure and capabilities of \yambo.
A more detailed description of how the code is actually utilized is presented in
Section~\ref{sec:components}, before some brief notes regarding installation 
in Section~\ref{sec:install}.
Finally, an illustrative example of a typical \yambo\ calculation is outlined
in Section~\ref{sec:lets_run_it}.

\section{Theoretical background}\label{sec:theo}

\subsection{Quasiparticles: the plasmon-pole approximation}\label{sec:QP}

MBPT is a rigorous approach 
based on the Green's function method, and
provides a proper framework for accurately computing excited state properties.
For details of the Green's function formalism and many-body techniques applied
to condensed matter, we refer the reader to several comprehensive papers in the
literature~\cite{MBPT_review,hedin}.
Here we shall just present some of the main equations used for the quasiparticle
and optical spectra calculations.  

\begin{figure} \centering
\epsfig{figure=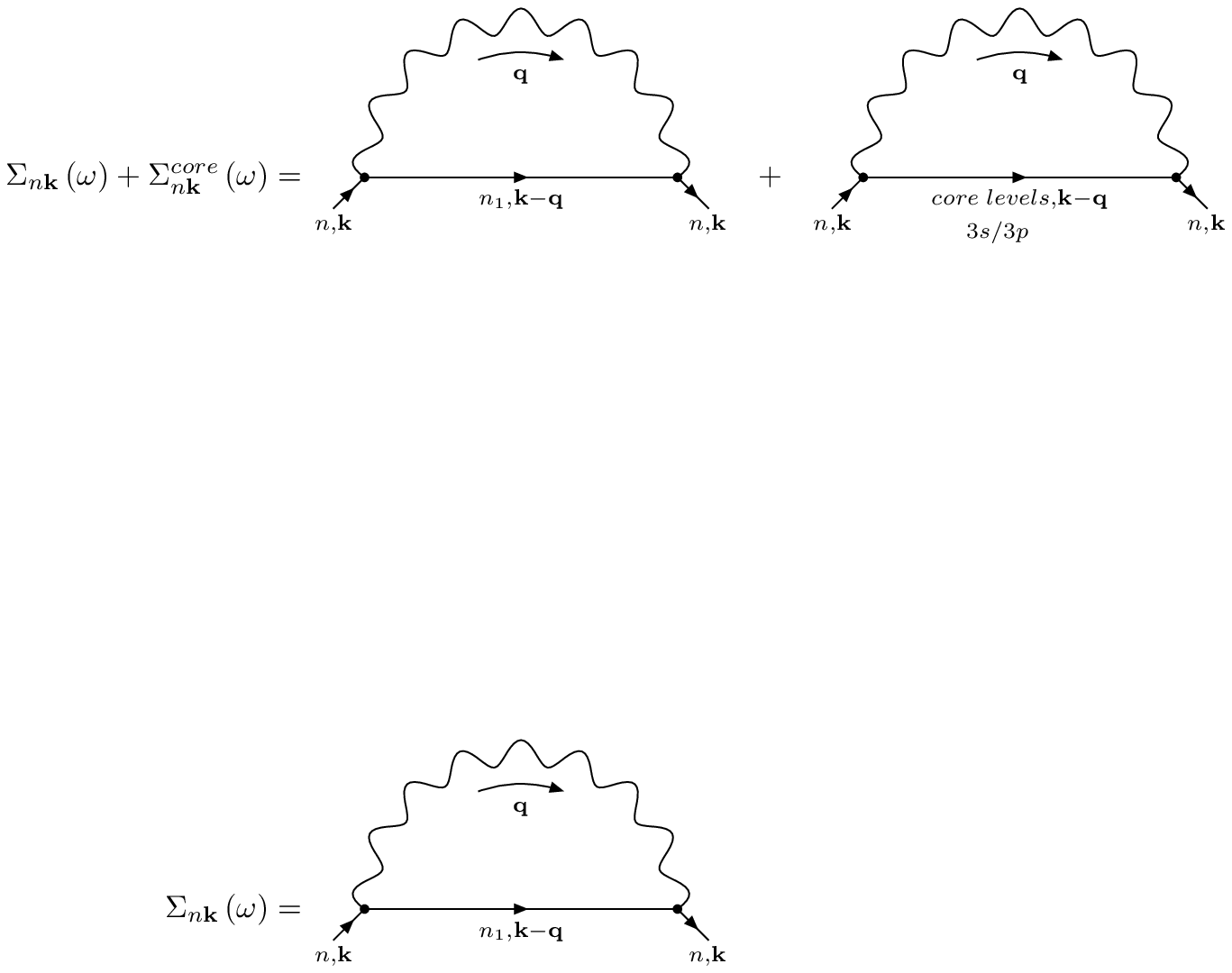,clip=,bbllx=130,bblly=425,bburx=330,bbury=520,width=.6\textwidth}
\caption{
Diagrammatic representation of the GW approximation for the self-energy 
operator.}
\label{fig:GW_diag}
\end{figure}

The basic component of a many-body perturbative expansion is the 
reference noninteracting
system, that in  \yambo\ is represented by the solution of the DFT
Kohn--Sham\,(KS) equations. 
In the following we will label these single particle levels as
$|n\kk\ra$, $n$ being the band index and $\kk$ the generic vector of the 
grid used to sample the Brillouin Zone\,(BZ). In this basis 
the noninteracting Green's function $G^0$ takes the form
\begin{align}
  G^0_{n\kk}\(\go\)=
  \frac{f_{n\kk}}{\go-\gee_{n\kk}-i0^+}+ \frac{1-f_{n\kk}}{\go-\gee_{n\kk}+i0^+},
  \label{eq:qp_g0}
\end{align}
$f_{n\kk}$ being the occupation factor and $\gee_{n\kk}$ the KS energies.
The basic relation between $G^0$ and the exact Green's function is given by the
Dyson equation
\begin{align}
  G_{n\kk}\(\go\)=\[\(G^0_{n\kk}\(\go\)\)^{-1}-\Sigma_{n\kk}\(\go\)+V^{\text{xc}}_{n\kk}\]^{-1},
  \label{eq:qp_dyson}
\end{align}
where the contribution due to the DFT exchange-correlation potential $V^{\text{xc}}_{n\kk}$ is removed
from the single-particle energies appearing in  $G^0_{n\kk}\(\go\)$ in order to prevent double
counting of correlation effects induced by the self-energy $\Sigma$.
Since the basic physical process that distinguishes a bare particle from a quasiparticle is
the screening of the particle by means of the polarization of the surrounding medium, 
\yambo\, uses the $GW$ approximation for the electronic self-energy $\Sigma$~\cite{GW_review}
which is diagrammatically depicted in Fig.~\ref{fig:GW_diag}. 
In this approximation the self-energy is a function of $G^0$ and of the 
inverse dynamical dielectric function $\epsilon^{-1}\(\rr_1,\rr_2;\go\)$, and it is composed of an exchange\,(x),
and  of a correlation\,(c) part,
\begin{align}
  \Sigma_{n\kk}\(\go\)=\Sigma^{x}_{n\kk}+\Sigma^{c}_{n\kk}\(\go\).
  \label{eq:qp_sig_sig_x_plus_sig_c}
\end{align}
The exchange part is simply the Fock term of the  Hartree--Fock self-energy, and it 
can be rewritten as
\begin{align}
 \Sigma^{x}_{n\kk}=\la n\kk|\Sigma^x(\rr_1,\rr_2)|n\kk\ra=
 -\sum_{m} \int_{BZ} \frac{d\qq}{(2\pi)^3}
  \sum_\GG v\(\qq+\GG\)
   \vert \rho_{nm}(\kk,\qq,\GG)\vert^2 f_{m(\kk-\qq)},
\label{eq:qp_sig_x}
\end{align}
where
$\rho_{nm}(\kk,\qq,\GG)=\la n\kk|e^{i(\qq+\GG)\cdot \rr)} |m\kk-\qq\ra$,
$\GG$ are the reciprocal lattice vectors, and $v\(\qq+\GG\)\equiv 4\pi/\vert \qq+\GG \vert^2$.
The correlation part of the self-energy is given by
\begin{multline}
 \Sigma^{c}_{n\kk}\(\go\)=\la n\kk|\Sigma^c(\rr_1,\rr_2;\go)|n\kk\ra=
  i\sum_{m} \int_{BZ} \frac{d\qq}{(2\pi)^3}
  \sum_{\GG,\GG'}\frac{4\pi}{\vert \qq+\GG \vert^2} \rho_{nm}(\kk,\qq,\GG) 
  \rho_{nm}(\kk,\qq,\GG') \times\\ 
  \int\,\d\go' G^{0}_{m \kk-\qq}\(\go-\go'\)
  \epsilon^{-1}_{\GG\GG'}\(\qq,\go'\).
\label{eq:qp_sig_c}
\end{multline}
The energy integral entering Eq.(\ref{eq:qp_sig_c}) can be solved once the
inverse dielectric function is known. The equation of motion for
 $\epsilon^{-1}$ follows from that of the reducible response function $\chi$~\cite{hedin}
as
\begin{align}
\epsilon^{-1}_{\GG\GG'}\(\qq,\go\)=\gd_{\GG\GG'}+v\(\qq+\GG\)\chi_{\GG\GG'}\(\qq,\go\).
\label{eq:gee_m1_vs_chi}
\end{align}
The $GW$ approximation for the self-energy is obtained when $\chi$ is calculated within the random phase
approximation\,(RPA)~\cite{hedin} 
\begin{align}
 \chi_{\GG\GG'}\(\qq,\go\)=\[\gd_{\GG\GG^{''}}-v(\qq+\GG^{''}) 
\chi^0_{\GG\GG^{''}}\(\qq,\go\)\]^{-1} \chi^0_{\GG^{''}\GG'}\(\qq,\go\).
\label{eq:dyson_for_chi}
\end{align}
The noninteracting response function is easily calculated in terms of the 
bare Green's function $G_0$:
\begin{multline}
\chi^0_{\GG\GG'}\(\qq,\go\)=2\sum_{nn'}\int_{BZ} \frac{d\kk}{(2\pi)^3}
\gr^*_{n'n\kk}\(\qq,\GG\)
\gr_{n'n\kk}\(\qq,\GG'\)f_{n\kk-\qq}\(1-f_{n'\kk}\)\times\\
\[\frac{1}{\go+\gee_{n\kk-\qq}-\gee_{n'\kk}+i0^+}-
\frac{1}{\go+\gee_{n'\kk}-\gee_{n\kk-\qq}-i0^+}\]. 
\label{eq:chi0}
\end{multline}
As the numerical integration of $\epsilon^{-1}$  in  Eq.(\ref{eq:qp_sig_c})  
would require the inversion of Eq.(\ref{eq:dyson_for_chi}) for many  
frequency points, \yambo\ adopts the plasmon-pole 
approximation\,(PPA) for the GW self-energy~\cite{GW_review}.
In the PPA the $\epsilon^{-1}$ function is approximated with a single pole function
\begin{align}
\epsilon^{-1}_{\GG\GG'}\(\qq,\go\)\approx \gd_{\GG\GG'}+
R_{\GG\GG'}\(\qq\)\[\(\go-\Omega_{\GG\GG'}\(\qq\)+i0^+\)^{-1}-
\(\go+\Omega_{\GG\GG'}\(\qq\)-i0^+\)^{-1}\],
\label{eq:PPA}
\end{align}
and the residuals $R_{\GG\GG'}$ and energies $\Omega_{\GG\GG'}$ are found by imposing the
PPA to reproduce the exact $\epsilon^{-1}$ function at $\go=0$ and $\go=iE_{\textrm{PPA}}$ ,
with $E_{\textrm{PPA}}$ being a suitable user-defined parameter.

Using Eq.~\ref{eq:qp_dyson} and assuming $f_{n\kk}$ to be either $1$ or $0$ we have that:
\begin{align}
  \(\go-\gee_{n\kk}\)  G_{n\kk}\(\go\) = 1+ \[\Sigma_{n\kk}\(\go\)-V^{\text{xc}}_{n\kk}\] G_{n\kk}\(\go\).
  \label{eq:d1}
\end{align}
The key approximation is now to take the first order Taylor expansion of the self-energy around 
$\gee_{n\kk}$ (Newton approximation) in order to get 
\begin{align}
  G_{n\kk}\(\go\)\approx
  Z_{n\kk}\[\frac{f_{n\kk}}{\go-E^{QP}_{n\kk}-i0^+}+ \frac{1-f_{n\kk}}{\go-E^{QP}_{n\kk}+i0^+}\],
  \label{eq:gqp}
\end{align}
with 
\begin{gather}
  E^{QP}_{n\kk}=\gee_{n\kk}+Z_{n\kk}\[\Sigma_{n\kk}\(\gee_{n\kk}\)-V^{\text{xc}}_{n\kk}\],
  \label{eq:qp1.a}\\
  Z_{n\kk}=\[1-\left.\frac{d\Sigma_{n\kk}\(\go\)}{d\go}\right|_{\go=\gee_{n\kk}}\]^{-1}.
  \label{eq:qp1.b}
\end{gather}
Eqs.~(\ref{eq:qp1.a}--\ref{eq:qp1.b}) constitute the QP approximation~\cite{GW_review}.

It is important to note, at this stage, that by including explicitly the electronic occupations
$f_{n\kk}$, \yambo\ can be equally applied to semiconductors,
insulators and metals. In the latter case, however, the plasmon-pole approximation can
be questionable when some of the valence orbitals are spatially localized
(like in {\it d} or {\it f} metals)~\cite{noPPAformetals}.  Nevertheless
for metals in general the RPA is an excellent approximation to the calculation
of optical properties, as the efficient screening occurring at the Fermi surface
prevents the formation of excitonic states.

\subsection{Optical properties: the Bethe--Salpeter equation}

The evaluation of the response function $\chi$ makes it possible to calculate the
macroscopic dynamical dielectric function $\epsilon_M$ and polarizability
$\alpha$. In particular the {\it macroscopic} dielectric function is defined
in terms of  the {\it microscopic} inverse  dielectric
function
as~\cite{ehrenreich}
\begin{align}
\epsilon_M\(\go\)\equiv\lim_{\qq\rightarrow 0}\frac{1}{\[\epsilon\(\qq,\go\)^{-1}\]_{\GG=0\,\GG'=0}},
\label{eq:epsilon}
\end{align}
where $\epsilon$ is the matrix in the space of reciprocal vectors $\GG$
defined in  Eq.~(\ref{eq:gee_m1_vs_chi}).
Equation~(\ref{eq:epsilon}) implies that, in general, one cannot take the simple spatial macroscopic
average of the dielectric function~\cite{ehrenreich} since the charge redistribution
induced by the interaction with light induces, in turn, the
formation of local microscopic fields---the local field effects. 
Such effects are particularly important in
nanoscale materials where confinement induces the formation of microscopic
fields that counteract the external applied perturbation~\cite{rmp}.
Similarly the dynamical polarizability of a zero-dimensional electronic
system is defined as
\begin{align}
\alpha\(\go\) = -\frac{\Omega}{4\pi}\lim_{\qq\rightarrow 0}\frac{1}{q^2}\chi_{\GG=0\,\GG'=0}\(\qq,\go\),
\label{eq:alpha}
\end{align}
where $\Omega$ is the unit cell  volume.

\begin{figure} \centering
\epsfig{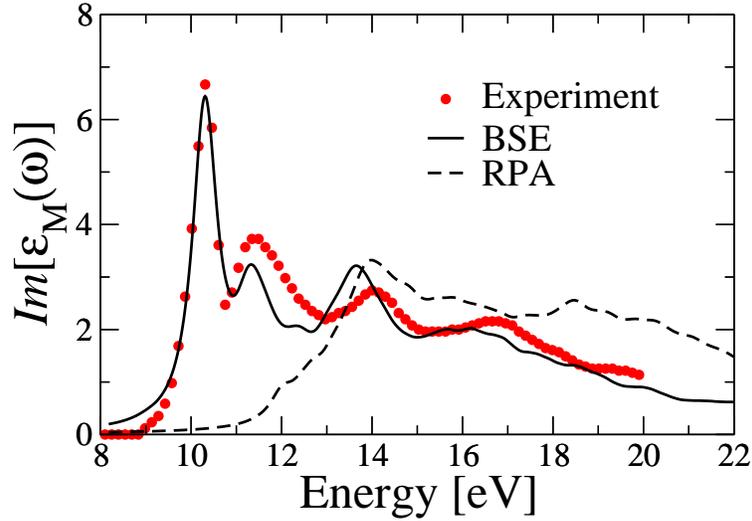}
\caption{The calculated absorption spectra of solid SiO$_2$ within the RPA and through solving 
the BS equation are compared with the experimental curve. Calculations taken from
Ref.~\cite{marini_fxc}, experiment from Ref.~\cite{sio2_exp}.
}
\label{fig:sio2}
\end{figure}

The RPA to Eqs.(\ref{eq:epsilon},\ref{eq:alpha}) is often inadequate to describe the electronic correlations 
occurring in the response function.
In practical applications the RPA 
does not yield optical absorption spectra in good agreement with experiments for
several insulating and metallic systems~\cite{rmp}.
For example, in the case of
SiO$_2$, this discrepancy has led to extensive debates over the past forty years
about the nature of the four well-defined peaks observed in the experiment (See Fig.~\ref{fig:sio2}).  
The reason for the poor performance of the RPA is that
the response function
$\chi$ measures the change in the electronic density induced by the external
applied potential. In a noninteracting system the RPA for $\chi$ is exact, but
self-energy corrections modify the electronic density 
and, consequently, the RPA approximation is not valid anymore.
Therefore, an estimation of the importance of corrections to the RPA can be
obtained by looking at the value of the gap correction. The larger the gap, 
the less 
adequate is the RPA. The gap of SiO$_2$, indeed, is $\sim$\,10\,eV and the RPA is not
even qualitatively correct. 

\begin{figure} \centering
\epsfig{figure=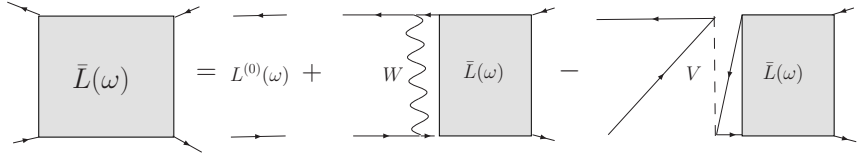,width=.7\textwidth}
\caption{
Diagrammatic representation of the BS equation.
}
\label{fig:BS_diag}
\end{figure}

The drawbacks of the RPA are solved by using a more elaborate equation of motion for
$\chi$ that takes into account the effect of electron--electron correlations. This 
is the BS equation~\cite{hedin} that can be introduced by using
the electron--hole\,(e--h) Green's function $L$. First we note that the  noninteracting  $\qq=0$
response function, Eq.(\ref{eq:chi0}), can be rewritten in terms
of the noninteracting  e--h Green's function $L^{0}$:
\begin{align}
\qlim\chi^0_{\GG\GG'}\(\qq,\go\)=-i\sum_{nn'\kk}
\lim_{\qq\rightarrow 0}\[\gr^*_{n'n\kk}\(\qq,\GG\)
\gr_{n'n\kk}\(\qq,\GG'\) \]
L^{0}_{nn'\kk}\(\go\).
\label{eq:chi0_as_L0}
\end{align}
To avoid the inversion of Eq.(\ref{eq:epsilon}), we
define a new interacting polarization~\cite{rmp} such that
$\epsilon_M\(\qq,\go\)\equiv1-v\(q\)\bar{\chi}_{\GG=0\,\GG'=0}\(\qq,\go\)$. This 
function defines a corresponding e--h Green's function  $\bar{L}$:
\begin{align}
\qlim\bar{\chi}_{\GG\GG'}\(\qq,\go\)=-i\sum_{nn'\kk}\sum_{mm'\kk'}
\qlim\[\gr^*_{n'n\kk}\(\qq,\GG\)
\gr_{m'm\kk'}\(\qq,\GG'\)\] \bar{L}_{\substack{nn'\kk\\mm'\kk'}}\(\go\).
\label{eq:chi_as_L}
\end{align}
The BS equation is an equation for $\bar{L}$, obtained by performing 
a second iteration of Hedin's equation~\cite{hedin}
\begin{align}
\bar{L}_{\substack{nn'\kk\\mm'\kk'}}\(\go\)=L^{0}_{nn'\kk}\(\go\)\[\gd_{nm}\gd_{n'm'}\gd_{\kk\kk'}
+i\sum_{ss'\kk_{1}} \Xi_{\substack{nn'\kk\\ss'\kk_{1}}}
\bar{L}_{\substack{ss'\kk_{1}\\mm'\kk'}}\(\go\)\].
\label{eq:BSE}
\end{align}
The BS equation naturally takes into account the electron--hole interaction in the response function as
shown schematically in Fig.~\ref{fig:BS_diag}.
The matrix 
$\Xi_{\substack{nn'\kk\\ss'\kk_{1}}}=W_{\substack{nn'\kk\\ss'\kk_{1}}}-2\bar{V}_{\substack{nn'\kk\\ss'\kk_{1}}}$ 
is the kernel of the 
BS equation composed of a direct electron--electron scattering term ($W$) plus an
exchange interaction ($\bar{V}$). Both $W$ and $\bar{V}$ are integrals of Bloch functions
\begin{gather}
W_{\substack{nn'\kk\\ss'\kk_{1}}}=\frac{1}{\Omega N_q}\sum_{\GG\GG'}
\gr_{ns}\(\kk,\qq=\kk-\kk_{1},\GG\) \gr_{n's'}^*\(\kk_{1},\qq=\kk-\kk_{1},\GG'\) \epsilon^{-1}_{\GG\GG'}v\(\qq+\GG'\),\\
\bar{V}_{\substack{nn'\kk\\ss'\kk_{1}}}=\frac{1}{\Omega N_q}\sum_{\GG\neq 0}
\gr_{nn'}\(\kk,\qq=0,\GG\) \gr_{ss'}^*\(\kk_{1},\qq=0,\GG\) v\(\GG\).
\label{eq:V_and_W}
\end{gather}
It is important to observe that the BS equation is a Dyson-like equation that should be inverted 
at each frequency point. The dimension of the $\bar{L}$ matrix can be, however, quite large
and the inversion of Eq.(\ref{eq:BSE}) practically impossible. However, 
for systems with a gap and at zero temperature the noninteracting e--h Green's function can be rewritten as
\begin{align}
\Re\[L^{0}_{nn'\kk}\(\go\)\]=i\frac{f_{n'\kk}-f_{n\kk}}{\go - \gee_{n\kk}+\gee_{n'\kk}}.
\label{eq:real_Lo}
\end{align}
As a consequence it can be shown~\cite{rmp} that the BS equation can be reduced to an eigenvalue problem of
the Hamiltonian $H$,
\begin{align}
H_{\substack{nn'\kk\\mm'\kk'}}=\(\gee_{n\kk}-\gee_{n'\kk}\)\gd_{nm}\gd_{n'm'}\gd_{\kk\kk'}
+\(f_{n'\kk}-f_{n\kk}\)\[2\bar{V}_{\substack{nn'\kk\\mm'\kk'}}-W_{\substack{nn'\kk\\mm'\kk'}}\].
\label{eq:BSE_H}
\end{align}
The Hamiltonian in Eq.~(\ref{eq:BSE_H}) is in general
non-Hermitian. Nevertheless \yambo\ adopts the standard
Tamm--Dancoff
approximation~\cite{FetterW71}, in which only e--h pairs at positive energy are
considered and the Hamiltonian $H$ is Hermitian. 
Finally the dielectric function can be expressed in terms of the eigenstates $|\gl\ra$
and eigenvalues $E_{\gl}$ of $H$:
\begin{align}
\epsilon_M\(\go\)\equiv 1-\lim_{\qq\rightarrow 0}\frac{8\pi}{|\qq|^2 \Omega N_q}
\sum_{nn'\kk}\sum_{mm'\kk'}
\gr^*_{n'n\kk}\(\qq,\GG\)
\gr_{m'm\kk'}\(\qq,\GG'\) 
\sum_{\gl} \frac{ A^{\gl}_{n'n\kk} \(A^{\gl}_{m'm\kk'}\)^*}{\go-E_{\gl}},
\label{eq:BSE_epsilon}
\end{align}
with $A^{\gl}_{n'n\kk}=\la n'n\kk | \gl \ra$ being the  eigenvectors of $H$.

In the case of semiconductors the BS approach
induces only a minor modification of the absorption spectrum. For wide-gap insulators, 
instead, the energies $E_{\gl}$ may fall within the single particle gap. In this case 
the eigenstates of the BS equation are called bound excitons~\cite{rmp}.
A typical example is shown in Fig.~\ref{fig:sio2} for the case of solid SiO$_2$~\cite{sio2,marini_fxc}.

\subsection{Time-dependent density functional theory}
Time-dependent density functional theory\,(TDDFT)~\cite{TDDFT_review} is gaining increasing popularity as 
an efficient tool for calculating
electronic excitations in finite systems, thanks to its simplicity and moderate computational cost.
In TDDFT the {\it exact} polarization function satisfies a Dyson-like equation that reads
\begin{align}
 \chi_{\GG\GG'}\(\qq,\go\)=
 \chi^{0}_{\GG\GG'}\(\qq,\go\)+\sum_{\GG^{''}}
 \[v(\qq+\GG^{''})+f_{\text{xc}}(\qq,\GG,\GG^{''})\] \chi_{\GG^{''}\GG'}\(\qq,\go\).
\label{eq:tddft_for_chi}
\end{align}
Eq.(\ref{eq:tddft_for_chi}) is similar to Eq.(\ref{eq:dyson_for_chi})
with an important addition, the xc kernel $f_{\text{xc}}$ that accounts for the exchange and correlation
effects. The exact form of the xc kernel 
is unknown, but the TDDFT
success relies also in the fact that even using the simplest adiabatic local density approximation\,(ALDA) 
for $f_{\text{xc}}$ a good accuracy in the evaluation of optical properties can be obtained~\cite{castro_08}.

We would like to mention here some common limitations of Eq.(\ref{eq:tddft_for_chi}), associated with
the use of plane-waves,  when 
applied to low-dimensional electronic systems.
As \yambo\ is a plane-wave code, it uses the super-cell approach to study small systems
like molecules and nanoscale materials.
This approach usually leads to two types of problems.

\begin{figure} \centering
\epsfig{figure=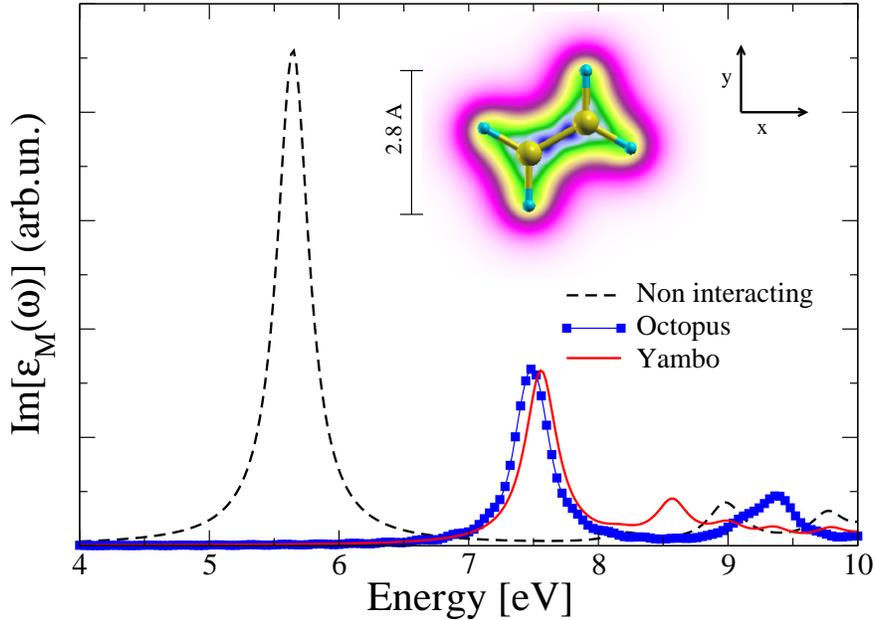,width=.7\textwidth}
\caption{Calculated photo-absorption spectra at the ALDA level along the $x$-direction of 
an ethylene
molecule using either the real-space real-time code {\tt Octopus}~\cite{octopus} 
and  \yambo\
in configuration space (see text for detail).
Inset: Contour plot of the electronic density of the ethylene molecule placed in a cubic super-cell of side 10\ \AA.  
}
\label{fig:alda}
\end{figure}

The first of these is the possibility of fictitious interactions
between super-cell replicas (images). To treat this problem,
\yambo\ is capable of removing the long-range tail of Coulomb potentials following the method
described in Ref.\cite{cutoff}.
The second problem 
derives from the numerical
instability of
Eq.(\ref{eq:tddft_for_chi}) that is induced by the presence of large regions of space with vanishing 
density, as occurs in large supercells (see inset of Fig.~\ref{fig:alda}).
Unfortunately, this vanishing density
induces some 
complications to Eq.(\ref{eq:tddft_for_chi}) when  $f_{\text{xc}}$ is evaluated at the ALDA
level.
Indeed, the ALDA kernel is defined by:
\begin{equation}
f_{\text{xc}}^{\text{ALDA}}(\rr,\rr';t)=\delta(\rr-\rr')\frac{dv_{xc}^{\text{HEG}}(n)}{dn}\vert_{n=n(\rr,t)},
\end{equation}  
where $v_{xc}^{\text{HEG}}(n)$ is the exchange and correlation potential of the homogeneous 
electron gas. 
In the region of space with vanishing density we have that $f_{\text{xc}}^{\text{HEG}}\(n\)\rightarrow\infty$
so that the evaluation of the term involving the $f_{\text{xc}}$ in Eq.~(\ref{eq:tddft_for_chi})
cannot be directly calculated in reciprocal space, because the  $f^{\text{ALDA}}_{\text{xc}}$
is not well defined.
To overcome this problem \yambo\  can solve the 
TDDFT equation within the ALDA in the basis of the e--h pairs, instead of
in reciprocal space [Eq.~(\ref{eq:tddft_for_chi})]. In this case the TDDFT equation has the same form of  the BS equation, Eq.~(\ref{eq:BSE}),
with the kernel $\Xi$  given by
\begin{align}
\Xi^{\text{ALDA}}_{\substack{nn'\kk\\ss'\kk'}}=-K^{\text{ALDA}}_{\substack{nn'\kk\\ss'\kk'}}-
\bar{V}_{\substack{nn'\kk\\ss'\kk'}},
\label{eq:alda_kernel}
\end{align}
with
\begin{equation}
\label{confalda}
K^{\text{ALDA}}_{\substack{nn'\kk\\mm'\kk'}}
=2\int\int d\rr d\rr' \phi^*_{n\kk}(\rr)\phi_{n'\kk}(\rr)
f_{\text{xc}}^{\text{ALDA}}(\rr,\rr';t=0)
\phi_{m'\kk'}(\rr')\phi^*_{m\kk'}(\rr').
\end{equation}
In Fig.~\ref{fig:alda} we show, as an example, the ALDA dynamical polarizability 
along the {\it x}-axis of 
the ethylene molecule.
The \yambo\ result is 
compared with a calculation made with the code {\tt octopus}\cite{octopus}
where all the quantities are defined in real-space, and the results are not affected
by the presence of regions with zero density and of images due to the super-cell approach.
We can see that \yambo\ well reproduces the main excitation peak at 7.5~eV, that
is composed by bound (localized) electron-hole states\cite{yVSo}.

\section{Some numerical aspects}\label{sec:numerics}
The \yambo\ code has been applied to a wide range of materials, from bulk compounds to
molecules and nanostructures. As a consequence we have developed several
{\it ad hoc} methods specifically designed to solve physical and numerical
problems that can be commonly encountered.

\subsection{The random integration method}\label{sec:RIM}
\begin{figure} \centering
\epsfig{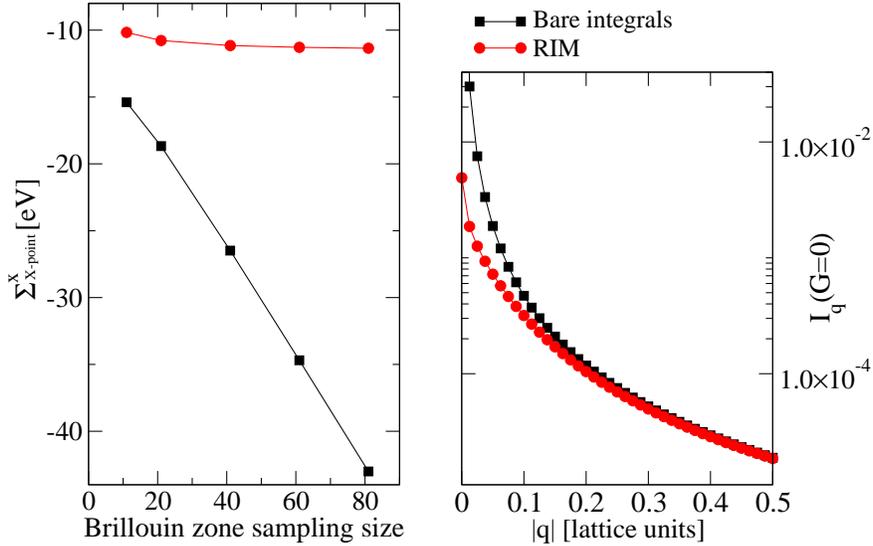}
\caption{Left panel: matrix element of $\Sigma_x$ at the X-point
of a {\it trans}-polyacetylene one-dimensional polymer~\cite{daniele_PA} in
function of the number of $\kk$ points used to sample the Brillouin zone, 
with and without the random integration method. We notice that the RIM
removes the divergence  of $\Sigma_x$ as a function of the size 
of the BZ sampling grid.
Right panel: $I_{\qq}\(\GG=0\)$ integrals as a function 
of $|\qq|$ for 40 $\kk$ Brillouin zone sampling size. 
The RIM cures the non analyticity of the Coulomb integral
near the $\Gamma$ point. Far from $\Gamma$ the RIM gradually  approaches the
bare integral.
}
\label{fig:RIM}
\end{figure}

The definition of the self-energy operator---and of many other quantities---requires  
an integration in the BZ. In practice this integral is 
replaced by some suitable grid of points.
Let's consider for simplicity only the Hartree--Fock self-energy,
given by Eq.(\ref{eq:qp_sig_x}). Using a finite grid of 
transferred momenta $\qq$ it reads
\begin{align}
 \Sigma^{x}_{n\kk} \approx
 -\frac{\(2\pi\)^3}{N_q\Omega}
 \sum_{m,\qq}
  \sum_\GG v\(\qq+\GG\)
   \vert \rho_{nm}(\kk,\qq,\GG)\vert^2 f_{m(\kk-\qq)}.
\label{eq:RIM_1}
\end{align}
In approximating $\Sigma^{x}_{n\kk}$ using a finite grid the key assumption
is that the integrand  of Eq.(\ref{eq:qp_sig_x}) must be a smooth function
of $\qq$. While this is generally true for the oscillators $\rho$ part,
this is not so trivial for the Coulomb potential that,
indeed, diverges for $\qq\rightarrow {\bf 0}$.

Nevertheless, in three-dimensional systems this divergence is not
affecting the calculation. In fact, the phase space volume associated
with the $\Gamma$ point reduces as $|\qq|^2$ when $\qq\rightarrow {\bf 0}$
and the divergence is {\it de facto} removed. In low-dimensional
systems this cancellation holds only if a
three-dimensional grid is used. When the Brillouin zone is sampled with 
lower-dimensional grid  instability problems appear in the evaluation 
of the Coulomb integral.
As shown in the right panel of Fig.~\ref{fig:RIM}, these instabilities may even
grow as a function of 
the Brillouin zone sampling size.
On the other hand using a three-dimensional grid is clearly
inconvenient,
and more importantly can make the
calculations extremely cumbersome. 

In order to avoid the use of a three-dimensional sampling in low-dimensional systems
\yambo\ offers two methods to remove the divergence arising from the 
Coulomb integrals. The first is based on a cutoff Coulomb technique which is discussed in
detail in Ref.\cite{cutoff}. The other is the so-called {\it random integration method}\,(RIM)~\cite{pulci}. 
In the RIM the HF self-energy is rewritten as
\begin{align}
 \Sigma^{x}_{n\kk}\approx
 -\sum_{m,\qq} 
  \vert \rho_{nm}(\kk,\qq,\GG)\vert^2 f_{m(\kk-\qq)}
  \int_{R_{\Gamma}} \frac{d\qq'}{(2\pi)^3}
  \sum_\GG v\(\qq+\qq'+\GG\).
\label{eq:RIM_2}
\end{align}
The integral of the Coulomb potential 
\begin{align}
  I_{\qq}\(\GG\)\equiv \int_{R_{\Gamma}} \frac{d\qq'}{(2\pi)^3}
  \sum_\GG v\(\qq+\qq'+\GG\)
\label{eq:RIM_3}
\end{align}
is evaluated in a region $R_{\Gamma}$ around the $\Gamma$ point.
Representing as $R_{\qq}$ the region $R_{\Gamma}$ translated in the general $\qq$ position, this
region is chosen in such a way that $R_{\Gamma}\cup R_{\qq_1}\cup R_{\qq_2}\dots R_{\qq_{N_q}}\equiv BZ$.
The RIM is based again on the uniformity {\it ansatz} described above, but restricted only
to the $\rho_{nm}$ factors.
The $I_{\qq}$ integrals are calculated via a three dimensional Monte Carlo technique~\cite{MC_integral}.

For practical purpose, it is necessary to evaluate  only $I_{\qq}\(\GG=0\)$ 
because far from the origin the approximation leading to Eq.~(\ref{eq:RIM_1}) 
is still accurate as shown in the right panel of Fig.~\ref{fig:RIM}.
The importance of the RIM is exemplified in the left panel of 
Fig.~(\ref{fig:RIM}) where we show the convergence of $\Sigma^x$ with respect to the $\kk$-point sampling
at the $X$ point of a quasi one-dimensional {\it trans}-polyacetylene polymer.

\subsection{The Lanczos--Haydock solver of the BS equation}\label{sec:haydock}

Once the BS (or the TDDFT) Hamiltonian $H$ [Eq.~(\ref{eq:BSE_H})]  has been
calculated in the basis of e--h pairs $|vc{\bf k}\ra$, \yambo\ offers two options for 
calculating the corresponding macroscopic dielectric function $\epsilon_M(\omega)$ and
related quantities (optical absorption, electron loss and dynamical
polarizability): either via the diagonalization, or 
via the Lanczos--Haydock (LH) recursion method~\cite{Haydock80}.
  
With the first option the program diagonalizes $H$
using the standard BLAS/LAPACK routines~\cite{Lapack} to find the
eigenvalues $E_{\lambda}$ and eigenstates $A^{\lambda}_{vc{\bf k}}$
that define the macroscopic dielectric function, Eq.(\ref{eq:BSE_epsilon}).
With the second option  Eq.~(\ref{eq:BSE_epsilon}) is
rewritten as, 
\begin{equation}\label{eq:epsm02}
\epsilon_M(\omega) = 1 - \langle P | (\omega - H)^{-1} | P \rangle,
\end{equation}
where $| P \rangle  = \lim_{{\bf q} \rightarrow 0}
\frac{1}{|{\bf q}|} |vc{\bf k}\rangle \langle {v{\bf k-q}} | e^{-i \qq\cdot\rr}
      |c{\bf k}\rangle $. 
Then Eq.~(\ref{eq:epsm02}) is calculated using the LH
method~\cite{Haydock80,Cini07}, a general algorithm to compute the matrix
elements of the Green's function $(\omega - H)^{-1}$ applied for the
first time to
solve the BS equation by Benedict and coworkers~\cite{BenedictS99}. 

The LH algorithm recursively builds an orthonormal basis
$ \{| q_i \rangle \}$ (Lanczos basis) in which $H$ is represented as a
real symmetric tridiagonal matrix,
\begin{equation}\label{eq:tridia}
T_j=\left(
\begin{matrix}
a_1    &  b_2   & 0      &  \cdots & 0 \\
b_2    &  a_2   & b_3    &       & \vdots \\
0      & \ddots & \ddots & \ddots & 0\\
\vdots &        & b_{j-1}& a_{j-1} & b_{j} \\
0      & \cdots & 0      & b_{j} & a_j \\
\end{matrix}
\right).
\end{equation}
The first vector $| q_1 \rangle$ of the Lanczos basis is set equal to $| P\rangle /
\|P\|$. The next vectors are calculated from the three-term relation
 \begin{equation}\label{eq:lnczos}
b_{j+1} |q_{j+1}\rangle = H |q_j\rangle - a_j | q_j\rangle -
b_j | q_{j-1}\rangle.   
\end{equation} 
In the Lanczos basis Eq.~(\ref{eq:epsm02}) becomes 
\begin{equation}\label{eq:cntfrc}
\epsilon_M^{j}(\omega) = 1 - \|P\|^2 \cfrac{1}{(\omega - a_1) - \cfrac{b_2^2}{(\omega - a_2) -
    \cfrac{b_3^2}{\dots}}}.  
\end{equation} 
At each iteration the program computes a new vector
$|q_{j+1}\rangle$ [Eq.~(\ref{eq:lnczos})], the matrix elements $a_j$, $b_{j+1}$, and thus a better approximation $\epsilon_M^{j}$   
for $\epsilon_M$ from Eq.~(\ref{eq:cntfrc}). The 
LH iteration procedure stops when the 
difference $|\epsilon_M^{j} -\epsilon_M^{(j-1)}|$ in
the given range of frequencies is smaller than a user-defined threshold.

\begin{figure} \centering
\epsfig{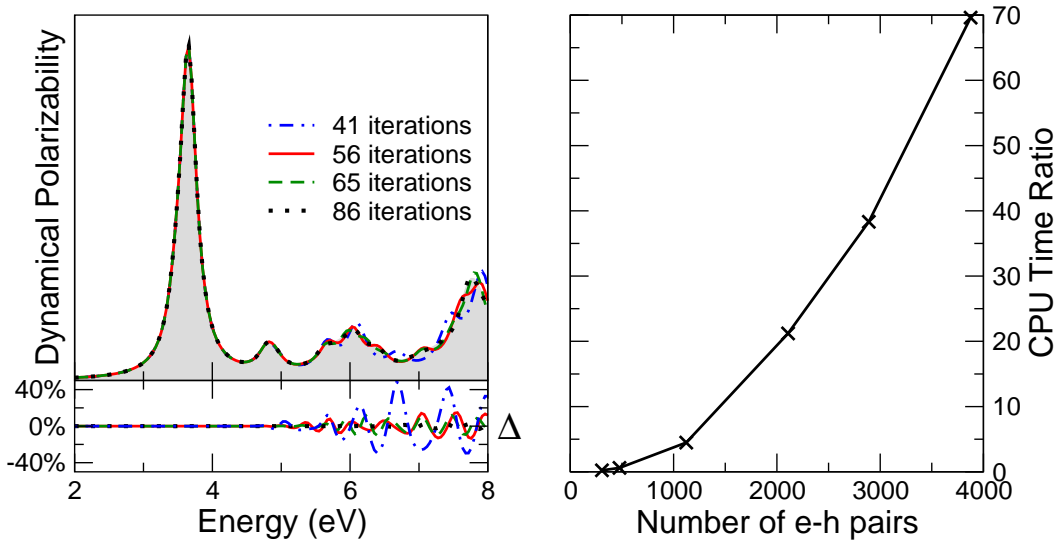}
\caption{Left top panel: Dynamical polarizability within the ALDA of the {\it
    trans}-azobenzene molecule calculated using either the diagonalization
  procedure (gray area) or the LH method varying the number of
  iterations. Left bottom panel: Relative error ($\Delta$) of LH
  method with respect to the diagonalization varying the number of
  iterations. Right panel: CPU time ratio between the diagonalization
  procedure and the LH method (50
  iterations) as a function of the
  e--h pairs included in the Hamiltonian.
}
\label{fig:LHtime}
\end{figure}

In practice when interested in a finite range of frequencies one
needs a number of iterations $k$ much smaller than the dimension $N$ of the
Hamiltonian (corresponding to number of e--h pairs included in
Eq.~\ref{eq:BSE_H}) to get an accurate $\epsilon_M(\omega)$. In the left panel
of Fig.~\ref{fig:LHtime} this is exemplified for the dynamical polarizability 
calculated within the ALDA of the {\it
  trans}-azobenzene molecule: $k\approx 50$ is enough for
getting an accurate spectrum up to 6 eV, with $N\approx 4000$. 
As a consequence, the LH method [$O(kN^2)$ floating point
operations] is usually much faster than the diagonalization
procedure [$O(N^3)$  floating point operations]. This is clearly shown
in the right panel of Fig.~\ref{fig:LHtime}: already for $N\approx
1000$ the LH method starts to
be more convenient, and for $N\approx 4000$ it is
about 70 times faster than the diagonalization procedure.
The LH method is also more convenient in terms of memory
since, in contrast with the diagonalization, only three vectors at a
time need to be stored. Furthermore, the whole procedure has
been efficiently parallelized in \yambo. 
The LH algorithm is, therefore, the recommended method in \yambo\
for the computation of the macroscopic dielectric function and related quantities, while the diagonalization should be used when
the eigenvalues and eigenvectors of $H$ are explicitly needed, as for
example for plotting the excitonic wavefunction via the
post-processing tool {\tt ypp}.

\section{Overview of the software}\label{sec:overview}

\begin{figure}\centering
  \epsfig{figure=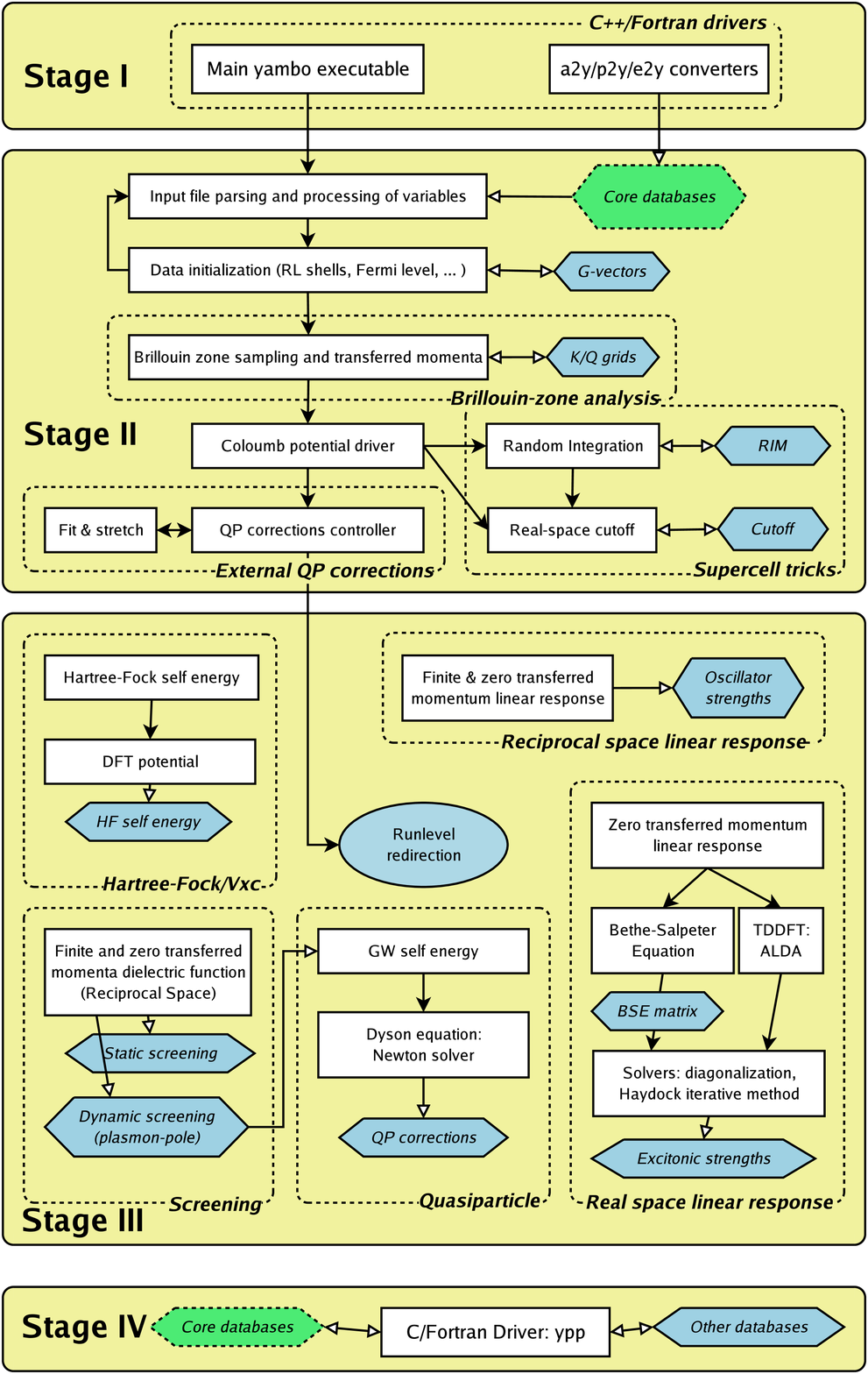,
          width=.9\textwidth}
  \caption{\label{fig:flowchart}
  Simplified schematic chart of the program. Diamond boxes denote some of the
  most important databases (see text).}
\end{figure}

The structure of the \yambo\ package can be separated into a number of stages, 
schematically depicted in Fig.~\ref{fig:flowchart}.
In the preliminary stage (I), a C/Fortran90 driver 
passes control to the main \yambo\ executable, or to the data converters
(\ay, \py, \ey). The purpose of the latter is to generate the core \emph{databases}
that contain the ground state data necessary for starting the code.
A mostly procedural data initialization stage (II) follows, where some general-purpose 
databases are prepared for later use.
The main physical calculations are performed in the third stage (III).
Finally, databases created by \yambo\ can be further manipulated using the post-processing tool 
\ypp\ in a fourth stage (IV).

Operation of the code follows a series of functionally distinct \emph{runlevels}. 
The main runlevels are activated 
by the user via the command-line interface (described in more detail in Sec.~\ref{sec:CLI}), 
and others are called automatically by the 
code where dependencies are present. Runlevels have a modular structure, in the sense that each 
one performs specific physical tasks and terminates (in most cases) with the creation of one or 
several databases written onto disk. These database files may then be accessed by different 
runlevels. Runlevels may be skipped in subsequent runs if a database is found that is compatible 
with the user requirements on execution. A more thorough discussion of the databases, and 
the \yambo\ I/O in general, can be found in Sec.~\ref{sec:IO}.

The main runlevels are now described in more detail.\\[.2cm]
Stage I:
\begin{enumerate}
\item C driver: governs the actions taken by the rest of the code. \yambo\ uses a standard
{\tt getopt} function~\cite{getopt} to parse the command line. This function is called
by all executables to acquire the user-defined options passed at the command line. 
The syntax is described briefly in Sec.~\ref{sec:CLI}.
\item Data import/converter: the ground-state electronic structure of the system to be studied is 
imported from ground state codes (see Sec.~\ref{sec:executables}), and converted into the core 
database files. 
\end{enumerate}
Stage II:
\begin{enumerate}
\item User input: if command line options were added in Stage I, the executable acts as an input 
file generator. On execution, the code reads default values for 
input parameters from existing databases and updates the values in the input file, after which the code terminates.
\item Data initialization: reorders $\GG$-vectors into spherical shells, calculates Fermi level and 
electronic occupations, sets up energy grids.
\item Brillouin-zone sampling: expands $\kk$-points to full BZ, generates $\qq$-point meshes, 
checks on uniformity of grids.
\end{enumerate}
Stage III:
\begin{enumerate}
\item Hartree-Fock/Vxc: calculates the matrix elements of the Hartree-Fock exchange self-energy and
of the DFT 
potential corresponding to the type of functional used in the ground state code run.
\item Screening: calculates static and dynamical inverse dielectric functions, for use in 
the evaluation of the GW self-energy and of the BS/TDDFT kernel.
\item Quasiparticle: calculates quasiparticle corrections to the
  Kohn--Sham band structure within the 
GW approximation.
\item Linear response: optical properties within RPA and ALDA with and without local field effects.
\item Bethe--Salpeter/TDDFT: creation of the BS/TDDFT Hamiltonian and subsequent diagonalization using 
LAPACK routines or via Lanczos--Haydock iterative procedure.
\end{enumerate}
Stage IV:
\begin{enumerate}
\item Post-processing: contains routines for creating $\kk$-point grids, analyzing single-particle
and excitonic 
wavefunctions and plotting the electronic wavefunctions and density.
\end{enumerate}

\section{Description of the individual software components}\label{sec:components}

\subsection{Main utilities}\label{sec:executables}

As described earlier, the \yambo\ package is in fact composed of three separate utilities.
The first of these is the set of converters (\ay, \py\ and \ey) which
generate the \yambo\ core databases from the output of other ground state codes.
The \ay\ converter imports data from the so-called  \texttt{KSS} file as generated
by the Abinit code\cite{abinit}, while
\py\ imports data written by the Quantum-ESPRESSO/PWscf\cite{pwscf} 
code in the \texttt{iotk} file format\cite{iotk}.
Alternatively, \ey\ can be used to import data from a netCDF formatted file 
written according to the ETSF file-format specifications\cite{etsf-nc}.
As a set of high-level libraries are available\cite{etsf-io} that are capable of reading and writing this
format, it is relatively easy to interface \yambo\ with other ground state codes.

The main computation tool is the \yambo\ executable itself. Depending on the command
line options used (see next section), it can act either as an input file generator or
as a straightforward serial or parallel executable.

The last utility is the \ypp\ post-processor, which is generally used to perform short
analysis of pre-calculated databases.

\subsection{Command line interface}\label{sec:CLI}

Thanks to the C/Fortran driver routine, \yambo\ offers a very user-friendly command-line 
interface for configuring the code at run-time and for creating and editing the main input file.
An extended list of options can be displayed by using the  \texttt{-H}
option (see Table~\ref{tab:synopsis}). 
\yambo\ options can be lowercase and uppercase. 
Upper case options may be 
added at run-time to specify input/output directories (\texttt{-I/-O}, see next Section), skip the MPI calls (\texttt{-N}), 
create a simple report of the existing database properties (\texttt{-D}), and so on. For instance, running 
\begin{verbatim}

yambo -N -I /scratch/tests/silicon-bulk -D

\end{verbatim}
will ask for a report of databases in the \texttt{/scratch/tests/silicon-bulk} directory, and 
force the parallel-compiled executable to run in serial.

Lower case options, instead, drive the input file editor. For example,
\begin{verbatim}

yambo -i -o c -I ~/tests/silicon-bulk

\end{verbatim}
will generate an input file (the default being \texttt{yambo.in}) that can be used to run the initialization steps 
(\texttt{-i}) and calculate linear response optics at the RPA level (\texttt{-o c}). A useful feature of the 
code is that default values for parameters are suggested based on their values in the most 
relevant database, if a value is not already present in the input file. Hence if 500 bands are 
written in the core databases, \yambo\ will suggest a range of 1--500 bands for calculating the 
noninteracting response function.

Since \yambo\ offers a huge range of tunable parameters, some of which are quite technical, it 
would be quite daunting for an average user to face the full list of parameters when running the 
code. Hence, an input-file verbosity flag ({\tt -V}) lets the user decide how detailed the input file is to 
appear (the remaining parameters are set to their default values). After checking the existing 
databases and input files, \yambo\ invoked in this way will redirect the user to edit the 
newly-created \texttt{yambo.in} file.

The C parser used by \yambo\ to process the input files is taken from the
{\tt octopus}~\cite{octopus} code.

\begin{table}\centering
\begin{tabular}{ll|ll|ll}  
 \hline 
\multicolumn{2}{c|}{\yambo} & \multicolumn{2}{c|}{\ay/\py/\ey} & \multicolumn{2}{c}{\ypp} \\
 \hline
 {\tt -h}           &Short Help &
 {\tt -h}           &Short Help &
 {\tt -h}           &Short Help  \\
 {\tt -H}           &Long Help & 
 {\tt -H}           &Long Help & 
 {\tt -H}           &Long Help \\
 {\tt -J <opt>}     &Job string identifier &
 {\tt -N      }     &Skip MPI initialization &
 {\tt -J <opt>}     &Job string identifier \\
 {\tt -V <int>}     &Input file verbosity &
 {\tt -F <opt>}     &Input file name/prefix &
 {\tt -F <opt>}     &Input file \\
 {\tt -F <opt>}     &Input file &
 {\tt -O <opt>}     &Output directory &
 {\tt -I <opt>}     &Core I/O directory \\
 {\tt -I <opt>}     &Core I/O directory &
 {\tt -S      }     &DataBases fragmentation &
 {\tt -O <opt>}     &Additional I/O directory \\
 {\tt -O <opt>}     &Additional I/O directory &
 {\tt -a <real>}    &Lattice constants factor &
 {\tt -C <opt> }    &Communications I/O directory \\
 {\tt -C <opt>}     &Communications I/O directory &
 {\tt -t      }     &Force no TR symmetry &
 {\tt -N      }     &Skip MPI initialization \\
 {\tt -N      }     &Skip MPI initialization &
 {\tt -y      }     &Force no symmetries &
 {\tt -S      }     &DataBases fragmentation \\
 {\tt -D      }     &DataBases properties &
 {\tt -w      }     &Force no wavefunctions &
 {\tt -k      }     &K-grid generator \\
 {\tt -S      }     &DataBases fragmentation &
                    & &
 {\tt -e <opt>}     &Excitons [(s)ort,(a)mplitude] \\
 {\tt -i      }     &Initialization &
                    & &
 {\tt -p <opt>}     &Plot [(e)xciton,(d)ensity,(w)aves] \\
 {\tt -o <opt>}     &Optics [opt=(c)hi/(b)se] &
                    & &
 {\tt -f}           &Free hole position [excitonic plot] \\
 {\tt -t <opt>}     &The TDDFTs [opt=(a)LDA/(l)RC] &
                    & &
                    & \\
 {\tt -c      }     &Coulomb interaction &
                    &&
                    & \\
 {\tt -x      }     &Hartree-Fock Self-energy and Vxc &
                    &&
                    & \\
 {\tt -b      }     &Static Inverse Dielectric Matrix &
                    &&
                    & \\
 {\tt -p <opt>}     &GW approximations [opt=(p)PA] &
                    &&
                    & \\
 {\tt -y <opt>}     &BS equation solver [opt=h/d] &
                    &&
                    & \\
 \hline 
\end{tabular}
\caption{
\footnotesize{Command line options for the various \yambo\ tools.}
\label{tab:synopsis}
}
\end{table}

\subsection{I/O: the \yambo\ databases}\label{sec:IO}

Files created and accessed by \yambo\ are classified according to their purpose and identified by 
a specific prefix:

\begin{enumerate}
\item Static database files (prefix \texttt{s.}, otherwise known as the core databases) 
are generated in Stage I (see Fig.~\ref{fig:flowchart}) 
by the \ay, \py\ and \ey\ converters.
They contain 
the information concerning the geometry, basis set, wavefunctions and energies, etc.
\item Stable database files (prefix \texttt{db.}) are created by \yambo\ at run-time, usually in 
Stage II. They are generally created once and used to store intermediate data designed to be re-read 
during subsequent executions.
\item Job-dependent database files (prefix \texttt{db.}) are also created at run-time but hold 
information that is usually specific to a particular runlevel (Stage III).
\item Output files (prefix \texttt{o.}) are usually created at the end of certain Stage III 
runlevels and are intended for human use (e.g. suitable for plotting or inspection).
\end{enumerate}

If the netCDF libraries~\cite{netcdf} have been configured for use, these database prefixes are further prefixed 
by \texttt{n}, e.g. \texttt{ndb.kindx} is the netCDF formatted version of \texttt{db.kindx}, the 
database containing the $\kk$/$\qq$-point grids. Users are strongly advised to use netCDF-linked 
executables, 
as the functionality of the code is enhanced.
In addition there are auxiliary run-time report files (prefix \texttt{r\_}) 
containing the run-time log information, while standard output can be redirected onto disk (prefix 
\texttt{l\_}).

Creation of each database is controlled by its own particular Fortran subroutine, e.g. 
\texttt{src/io/ioQP.F} creates the database of QP corrections \texttt{db.QP}. However, the actual 
writing of data is performed at a low level by some common modules, in order to minimize problems 
associated with portability. Some of the most important databases and their respective runlevels 
are indicated in Fig.~\ref{fig:flowchart} as diamond--shaped boxes.

In order to treat systems with large memory or disk requirements \yambo\ offers the 
capability of fragmenting the larger databases into chunks (\texttt{-S}). This functionality is most 
frequently utilized in the splitting of the ground-state wavefunction files according to $\kk$-points 
and/or bands, and in the division of the Bethe--Salpeter Hamiltonian according to the {\kk}-point 
index.

By default, all databases are stored in the subfolder \texttt{./SAVE} of the
working directory. More specific control of 
I/O directories may be accessed through the various uppercase command line options. For instance, 
it is common to store the static databases in a `core' directory (\yambo\ \texttt{-I COREPATH}), so that 
they can be shared by different processes. Dynamically-created databases can then be placed 
elsewhere (\yambo\ \texttt{-O OTHER}). Furthermore, users may organize their work better according to a 
``job string" identifier (\yambo\ \texttt{-J JOBNAME}), so that databases and output files are placed in a 
subdirectory with a name specified by the user. Finally, long jobs, like the creation or 
diagonalization of the Bethe--Salpeter Hamiltonian, place intermediate results in a \texttt{RESTART} folder, 
so that the job can be continued following a crash or if cpu time limits are exceeded.

\section{Installation instructions}\label{sec:install}

\yambo\ makes use of the GNU autotools suite (automake/autoconf/libtool) for installation. 
Hence, the standard procedure of
\begin{verbatim}

./configure
make all

\end{verbatim}
should create executables for \yambo, the \ypp\ postprocessor and the basic Abinit converter \ay\,
and place them in the \texttt{bin/} folder of the \yambo\ source directory.
Detailed installation instructions are available in the manual, including how to enable the 
\py\ and \ey\ converters, how to link with netCDF, BLACS and FFTW libraries, change installation
options, and so on.
In short, a somewhat complete list of compile-time options can be inspected by running the standard
command
\begin{verbatim}

./configure --help
\end{verbatim}

\section{Running \yambo: excitonic effects in bulk silicon}\label{sec:lets_run_it}

In this section we outline the basic steps involved in a typical production run of \yambo.
Normally one would of course start by creating the core databases by importing the
output data of some ground state code (see Sec.~\ref{sec:executables}).
For the purpose of illustration, however, we have included in the \yambo\ source a pre-compiled 
set of core databases for bulk silicon in the \texttt{doc/sample} directory (these netCDF
formatted databases can be
extracted by following the instructions in the \texttt{doc/sample/bulk\_silicon/README} file).

Once the core databases (\texttt{ns.db1, ns.wf}) are extracted, running
\yambo\ in the \texttt{doc/sample/bulk\_silicon/} directory
produces as standard output:
{\small
\begin{verbatim}
> yambo 
 
 <---> [01] Job Setup
 <---> [02] Input variables setup
 <---> [02.01] K-grid lattice
 <---> [02.02] RL shells
 <---> Shells finder |####################| [100%] --(E) --(X)
 <---> [02.03] Input (E)nergies[ev] & Occupations
 <---> [03] Transferred momenta grid
 <---> X indexes |####################| [100%] --(E) --(X)
 <---> SE indexes |####################| [100%] --(E) --(X)
 <---> [04] Game Over & Game summary

\end{verbatim}
}
This is the Stage II run that \yambo\ enters by default whenever, as in
this case, no input file has been specified.
The results of this setup run (BZ sampling indexes, \GG\ vector shells, etc.) are stored 
in two new database files, {\tt SAVE/ndb.gops} and {\tt SAVE/ndb.kindx}, that are to be
re-used in subsequent runs.
Besides the two databases, \yambo\ writes a report file, \texttt{r\_setup}, that contains a detailed list
of information about the run and the system. Every runlevel generates an appropriate report file.

At this stage we can carry out a more interesting calculation:
as an example we will calculate the optical absorption spectra of bulk
silicon including excitonic effects. To solve the BS equation we will 
employ the LH algorithm described in Sec.~\ref{sec:haydock}.  
First we create the input with the command-line based user interface.  
From Table~\ref{tab:synopsis} we see that the options to use are 
\begin{verbatim}

> yambo -b -o b -y h

\end{verbatim}
This command creates the {\tt yambo.in} input file
shown in Table~\ref{tab:input_file} and opens it in the default text editor, {\tt vi}.
\begin{table}\centering
\begin{tabular}{ll}  
 \hline 
optics              & \# [R OPT] Optics \\
bse                 &        \# [R BSK] Bethe Salpeter Equation. \\
em1s              &          \# [R Xs] Static Inverse Dielectric Matrix \\
bss               &          \# [R BSS] Bethe Salpeter Equation solver \\
BSresKmod= "xc"    &         \# [BSK] Resonant Kernel mode. (`x`;`c`;`d`) \\
\% BSEBands & \\
   1 {\tt |}  50  {\tt |}          &       \# [BSK] Bands range\\
\%  & \\
BSENGBlk= 1              RL &\# [BSK] Screened interaction block size \\
BSENGexx=  411           RL &\# [BSK] Exchange components \\
\% QpntsRXs  & \\ 
  1 {\tt |} 19 {\tt |}     &               \# [Xs] Transferred momenta \\
\% & \\
\% BndsRnXs & \\
   1 {\tt |}  50 {\tt |}         &        \# [Xs] Polarization function bands \\
\% & \\
NGsBlkXs= 1              RL &  \# [Xs] Response block size \\
\% LongDrXs & \\
 1.000000 {\tt |} 0.000000 {\tt |} 0.000000 |   &  \# [Xs] [cc] Electric Field \\
\%  & \\
BSSmod= "h"           &      \# [BSS] Solvers `h/d/i/t` \\
\% BEnRange & \\
  0.00000 {\tt |} 10.00000 {\tt |}   eV  & \# [BSS] Energy range \\
\% & \\
\% BDmRange & \\
  0.10000 {\tt |}  0.10000 {\tt |}   eV &  \# [BSS] Damping range \\
\% & \\
BEnSteps= 100           &    \# [BSS] Energy steps \\
\% BLongDir & \\
1.000000 {\tt |} 0.000000 {\tt |} 0.000000 {\tt |} &    \# [BSS] [cc] Electric Field \\
\% & \\
 \hline 
\end{tabular}
\caption{\footnotesize{\yambo\ input file needed to perform a calculation of the
optical properties of bulk silicon including excitonic effects (see text).}}
\label{tab:input_file}
\end{table}
Note that, by reading the databases generated in the setup run, \yambo\ already knows that
there are 19 momenta permitted by the BZ sampling, and that 
 50 bands were calculated in the preceding ground-state run.
To perform a test calculation we change some of the values in the input.
\begin{itemize}
\item In Eq.(\ref{eq:chi_as_L}), we restrict the summation 
  from band $2$ to band $6$: 
\begin{verbatim}
% BSEBands
   2 |  6 |                 # [BSK] Bands range
%
\end{verbatim}
\item In the statically screened interaction we use just $51$ RL vectors: 
\begin{verbatim}
BSENGBlk= 51             RL  # [BSK] Screened interaction block size
NGsBlkXs= 51             RL  # [Xs] Response block size
\end{verbatim}
\item  As the LH method is very fast we increase the number of points on the frequency axis
by setting {\tt BEnSteps=1000}. Moreover, to mimic the experimental width of
the E$_1$ and E$_2$ peaks, we set
\begin{verbatim}
% BDmRange
  0.02000 |  0.8000 |   eV  # [BSS] Damping range
%
\end{verbatim}
\end{itemize}
Calling \yambo\ again, without command line options, will start the calculation,
which should last for some minutes. 
The progress of the calculation (including expected/elapsed time for
each time-consuming operation) can be followed from the standard
output, or in case of background execution, read from the {\tt
  l\_optics\_bse\_em1s\_bss} log file.  

At the end of the run several new files appear in the {\tt SAVE}
folder:   
the dielectric function {\tt ndb.em1s}, the BS Hamiltonian matrix {\tt ndb.BS\_Q1},  and 
the information needed to restart the LH iterative
procedure in {\tt ndb.Haydock\_restart}. 
Furthermore, the run generates
two files in the working directory, the report file ({\tt r\_optics\_bse\_em1s\_bss}) and
the output file ({\tt o.eps\_q001-bh}).

\begin{figure} \centering
\epsfig{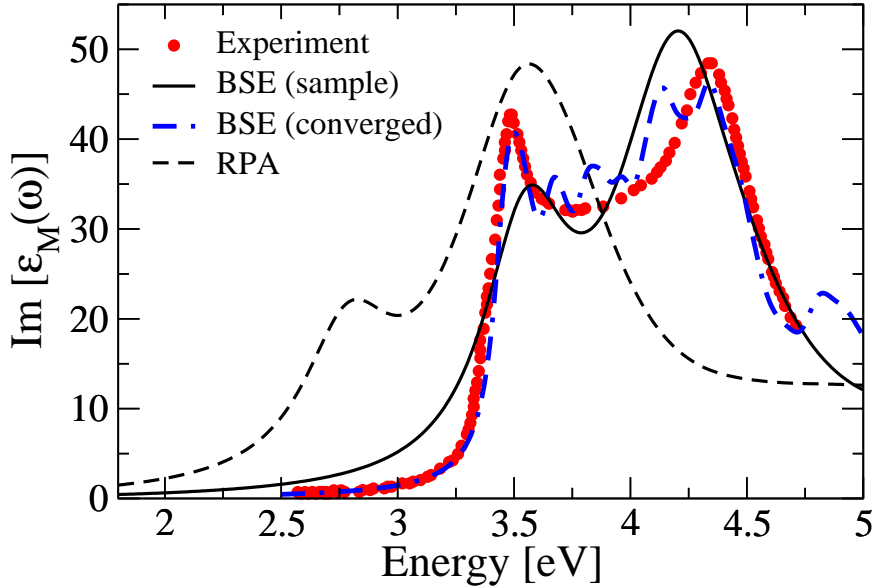}
\caption{Optical spectrum of bulk silicon.
The result of the sample run described in the text is compared with
a more converged calculation~\cite{convergedSi} and with the
experiment~\cite{jellison} .
Note that the result
of the sample run has been red--shifted by 1\,eV to simulate the
correct QP gap that, instead, is calculated 
in the converged spectrum~\cite{convergedSi}.}
\label{fig:bulk_si}
\end{figure}

The output file contains the calculated
spectra (BS and RPA spectra) that can be
visualized without further processing with standard plotting tools.
Plotting the second versus the first column of the {\tt o.eps\_q001-bh} file
gives the well-known absorption spectrum 
of bulk silicon within BSE. The result of the run is shown in Fig.~\ref{fig:bulk_si}, where
it is compared with the independent particle (RPA) spectrum (fourth versus the first column of the same
file), with the experimental spectrum~\cite{jellison} (included in the 
{\tt doc/sample/bulk\_silicon/experiment.dat} file) and with the result of a more converged 
calculation~\cite{convergedSi}.

\section{Appendices}\label{sec:appendices}

\subsection{The $\chi^0$ poles accumulation}\label{sec:Xo_grid}
The noninteracting response function $\chi^0$ is a key ingredient
of the \yambo\ code. It enters, for example, in the definition of the RPA
dielectric function and of the GW self-energy.
Nevertheless the practical evaluation of $\chi^{0}$ is often a
bottleneck in realistic calculations. The reason is that, as shown in Eq.(\ref{eq:chi0}),
$\chi^{0}$ contains a triple summation over the $\kk$ points and the occupied and empty electronic
levels.
For systems with  a large number of electrons this summation can easily reach 
millions of elements, that must be multiplied by the number of 
frequencies. 

To provide a tunable and efficient tool to reduce the numerical effort in evaluating 
$\chi^{0}$  \yambo\ follows the general idea of the algorithm proposed by
Miyake~\cite{miyake} to rewrite $\chi^{0}$  as
\begin{align}
\chi^0_{\GG\GG'}\(\qq,\go\)=
\frac{2}{\Omega N_k}
\sum_{p}
\[
\sum_{nn'\kk\in I_p}
\gr^*_{n'n\kk}\(\qq,\GG\)
\gr_{n'n\kk}\(\qq,\GG'\)
\]
F_p
\[\frac{1}{\go+E_p+i0^+}-
\frac{1}{\go-E_p-i0^+}\].
\label{eq:chi0_coarse}
\end{align}
In Eq.~(\ref{eq:chi0_coarse}) the $I_p$ are groups of e--h indexes
($nn'\kk$) with similar 
e--h energies $\gee_{n'\kk}-\gee_{n\kk-\qq}$. The latter are approximated with 
a single pole at $E_p$  with occupation $F_p$.
The number of groups created can be controlled by the user by tuning the
{\tt CGrdSp} variable in the input file. 

The important difference with  Eq.(\ref{eq:chi0}) is that, in
Eq.(\ref{eq:chi0_coarse}) the evaluation of the oscillators is partially decoupled
from the energy dependence. As a consequence by decreasing the groups of 
e--h pairs it is possible to make the evaluation of $\chi^0$ almost
independent on the number of frequencies.

\subsection{Oscillator symmetries}\label{sec:BSE_syms}
The oscillators 
\begin{align}\label{eq:BSE_syms_1}
\rho_{nm}(\kk,\qq,\GG)=\la n\kk|e^{i\(\qq+\GG\)\cdot \rr} |m\kk-\qq\ra=
\int_{\Omega}\,d\rr\, \psi^*_{n\kk}\(\rr\)e^{i\(\qq+\GG\)\cdot \rr}  \psi_{n\kk-\qq}\(\rr\)
\end{align}
appear in almost all the quantities calculated in \yambo.  They are evaluated
using efficient Fast Fourier Techniques\,(FFT)~\cite{goedecker}. Nevertheless \yambo\ uses
symmetry arguments to reduce the number of calls to the FFT interface. 

If we specify the symmetry operation by rewriting a general point in the BZ
as $\kk=R\kk_{IBZ}$, with $\kk_{IBZ}$ defined in the irreducible wedge of the BZ,
Eq.(\ref{eq:BSE_syms_1}) reads
\begin{align}\label{eq:BSE_syms_2}
\rho_{nm}(\kk,\qq,\GG)=
\int_{\Omega}\,d\rr 
\[e^{-iR\kk_{IBZ}} u^*_{nR\kk_{IBZ}}\(\rr\)\]
e^{i\(\qq+\GG\)\cdot \rr} 
\[e^{iR'\(\kk-\qq\)_{IBZ}-i\GG_0\cdot\rr}u^*_{mR'\(\kk-\qq\)_{IBZ}}\(\rr\)\],
\end{align}
with $\GG_0$ defined as $\GG_0=\kk-\qq-\(\kk-\qq\)_{IBZ}$. Hence we find that
\begin{align}\label{eq:BSE_syms_3}
\rho_{nm}(\kk,\qq,\GG)=
\int_{\Omega}\,d\rr\, 
u^*_{nR\kk_{IBZ}}\(\rr\)
e^{i\(\GG-\GG_0\)\cdot \rr} 
u_{mR'\(\kk-\qq\)_{IBZ}}\(\rr\).
\end{align}
In case of spatial symmetries (a similar procedure applies to the time-reversal
symmetry)
we can rewrite the left side rotated wavefunction as $u^*_{nR\kk_{IBZ}}\(\rr\)=
u^*_{n\kk_{IBZ}}\(R^{-1}\rr\)$, so that
\begin{align}\label{eq:BSE_syms_4}
\rho_{nm}(\kk,\qq,\GG)=
\int_{\Omega}\,d\rr\, 
u^*_{n\kk_{IBZ}}\(\rr\)
e^{iR^{-1}\(\GG-\GG_0\)\cdot \rr} 
u_{mR^{-1}R'\(\kk-\qq\)_{IBZ}}\(\rr\).
\end{align}
Finally we notice that the symmetries constitute a group, and, consequently,
$R^{-1}R'=S$, with $S$ a symmetry operation.
Thus, at difference with Eq.(\ref{eq:BSE_syms_1}), Eq.(\ref{eq:BSE_syms_4}) depends
only on one symmetry index. By using  Eq.(\ref{eq:BSE_syms_4})
the computational cost of calculating all the oscillators at a given $\kk$ point is reduced by
the number of symmetries in the star of $\kk$.
 
\section{Acknowledgments}
This work was partially supported by the FP6 European Network of
Excellence Nanoquanta (NMP4-CT-2004-500198).
We acknowledge grant support for code development and testing from the 
CINECA (account \texttt{cne2fm2h}) and CASPUR supercomputing centres.
We would like to thank  Rodolfo Del Sole and Angel Rubio for their active and
continuing support towards 
the development of the code, both as \yambo\ and during the time 
the code was known as {\tt SELF}.

We would also like to thank the people that have contributed in some way
to the development of the code: C. Attaccalite, M. Palummo, M. Gatti, F. De Fausti, 
M. Bockstedte, L. Wirtz,
G. Onida,
X. Gonze.


\begin{thebibliography}{100}
\bibitem{yVSo}
 The high 
 energy peaks appearing in Fig.~\ref{fig:alda} correspond to the excitation of unlocalized orbitals whose different 
 description in the \yambo\ and {\tt octopus} codes leads to the discrepancies observed in the
 figure. Nevertheless,
 these differences can be systematically removed by increasing the dimension of 
 the basis size.
\end{thebibliography}
\end{document}